\documentclass[%
 reprint,
superscriptaddress,
 amsmath,amssymb,
 aps,
prb,
floatfix,
]{revtex4-2}
\usepackage{xcolor}
\usepackage{graphicx}
\usepackage{dcolumn}
\usepackage{mathtools}

\begin{document}

\title{
Plasmonic lattice Kerker effect in UV-Vis spectral range
}

\author{V.S. Gerasimov}
\email{gerasimov@icm.krasn.ru}

\author{A.E.~Ershov}
  \affiliation{Siberian Federal University, Krasnoyarsk, 660041, Russia}
  \affiliation{Institute of Computational Modelling of the Siberian Branch of the Russian Academy of Sciences, Krasnoyarsk, 660036, Russia}
\author{R.G.~Bikbaev}
  \affiliation{Siberian Federal University, Krasnoyarsk, 660041, Russia}
  \affiliation{L.~V.~Kirensky Institute of Physics, Federal Research Center KSC SB RAS, Krasnoyarsk, 660036, Russia}
\author{I.L.~Rasskazov}
  \affiliation{The Institute of Optics, University of Rochester, Rochester, New York 14627, United States}
\author{I.L.~Isaev}
   \affiliation{Institute of Computational Modelling of the Siberian Branch of the Russian Academy of Sciences, Krasnoyarsk, 660036, Russia}
\author{P.N.~Semina}
  \affiliation{Siberian Federal University, Krasnoyarsk, 660041, Russia}
\author{A.S.~Kostyukov}
  \affiliation{Siberian Federal University, Krasnoyarsk, 660041, Russia}
\author{V.I.~Zakomirnyi}
  \affiliation{Siberian Federal University, Krasnoyarsk, 660041, Russia}
  \affiliation{Institute of Computational Modelling of the Siberian Branch of the Russian Academy of Sciences, Krasnoyarsk, 660036, Russia}
\author{S.P.~Polyutov}
    \email{spolyutov@sfu-kras.ru}
    \affiliation{Siberian Federal University, Krasnoyarsk, 660041, Russia}
\author{S.V.~Karpov}
   \affiliation{Siberian Federal University, Krasnoyarsk, 660041, Russia}  \affiliation{L.~V.~Kirensky Institute of Physics, Federal Research Center KSC SB RAS, Krasnoyarsk, 660036, Russia}

\date{\today}

\begin{abstract}
Mostly forsaken, but revived after the emergence of all-dielectric nanophotonics, the Kerker effect can be observed in a variety of nanostructures from high-index constituents with strong electric and magnetic Mie resonances. 
Necessary requirement for the existence of a magnetic response limits the use of generally non-magnetic conventional plasmonic nanostructures for the Kerker effect.
In spite of this, we demonstrate here for the first time the emergence of the lattice Kerker effect in regular plasmonic Al nanostructures.
Collective lattice oscillations emerging from delicate interplay between Rayleigh anomalies and localized surface plasmon resonances both of electric and magnetic dipoles, and electric and magnetic quadrupoles result in suppression of the backscattering in a broad spectral range.
Variation of geometrical parameters of Al arrays allows for tailoring lattice Kerker effect throughout UV and visible wavelength ranges, which is close to impossible to achieve using other plasmonic or all-dielectric materials. 
It is argued that our results set the ground for wide ramifications in the plasmonics and further application of the Kerker effect. 

\end{abstract}

\maketitle
\section{Introduction}
The concept of backscattering suppression of light by a single spherical particle has been proposed over three decades ago by Kerker et al.~\cite{Kerker:83}
The essential pre-requisite for this effect is the equivalence of permittivity and permeability of a sphere, $\varepsilon=\mu$, which implies a co-existence of pronounced electric and magnetic responses at the same frequency.
This exciting idea did not receive a lot of attention in the past because such materials are close to impossible to find in nature.
However, the situation has changed dramatically with the emergence of all-dielectric nanophotonics~\cite{Staude2019}.
Optically induced magnetic moments mediate so-called ``artificial magnetism''~\cite{Evlyukhin2012}, which surpasses the requirement for the material to exhibit conventional magnetic properties.
Thus, experimental evidence of the Kerker effect has been provided for single high-index Si~\cite{Fu2013} or GaAs~\cite{Person2013} nanoparticles (NPs).
After these pioneering experimental observations, the Kerker effect has been demonstrated in numerous setups~\cite{Staude2013,Alaee2015,Shen2016,Terekhov2017,Wiecha2017,Yang2017,Liu2019,Barhom2019,Pan2019,Shamkhi2019,Liu2020} with a promising applications in a variety of endeavors such as sensing~\cite{Bag2018}, imaging~\cite{Dubois2018} and others~\cite{Liu2018}.
These advances of all-dielectric nanophotonics left plasmonics much in a shadow. Inherently weak magnetic response of metal NPs, together with losses makes it difficult~\cite{Olmos-Trigo2020} yet possible~\cite{Yang2020e} to harness the Kerker effect in single plasmonic NPs.
Moreover, different combinations of plasmonic materials with all-dielectric~\cite{Poutrina2013,Pors2015} or magnetic~\cite{Kataja2016} structures may satisfy the Kerker condition.

On a larger scale, i.e., in arrays of NPs, the Kerker effect can be implemented via collective lattice resonances (CLRs)~\cite{Ross2016,Kravets2018}.
CLRs emerging in arrays of both plasmonic~\cite{Zou2004a,Markel2005,Auguie2008,Kravets2008} and all-dielectric~\cite{Evlyukhin2010a,Li2018a,Utyushev2020} NPs are high-quality modes originating from the coupling between Rayleigh anomaly and Mie resonances of a single NP.
By tailoring the configuration of the array and the shape of constituents, the \textit{lattice} Kerker effect can be observed.
Lattice Kerker effect arises due to interaction between lattice modes (i.e., CLRs) and resonances in a single NP, while conventional Kerker effect is based on resonances in a single NP.
We emphasize that the lattice Kerker effect is heavily studied in \textit{all-dielectric} nanostructures with strong magnetic dipole (or quadrupole) resonances~\cite{Babicheva2017,Babicheva2018,Babicheva2018a,Babicheva2018b} with only one exception of arrays of relatively large \textit{plasmonic} Au NPs exhibiting Kerker effect at 750~nm~\cite{Babicheva2018c}.
Rapidly developing aluminum plasmonics~\cite{Maidecchi2013,Knight2014,Gerard2015} provides a solid ground for CLRs~\cite{Yang2016,Khlopin2017,Esposito2019,Kawachiya2019,Zhu2020}, but most of the studies are traditionally limited to purely electric interactions, either dipole or dipole-quadrupole~\cite{Yang2016}.
In our recent work~\cite{Ershov2020}, we have shown that plasmonic arrays of Al NPs, very much similar to all-dielectric nanostructures, support magnetic dipole or quadrupole CLRs. 
We take the advantage of this property and for the first time demonstrate the Kerker effect in \textit{plasmonic} Al metasurfaces.
It is worth to notice that Al is the only plasmonic material which manifests localized surface plasmon resonance in UV wavelength range.

\begin{figure*}[ht!]
 \centering
 \includegraphics[width=6.5in]{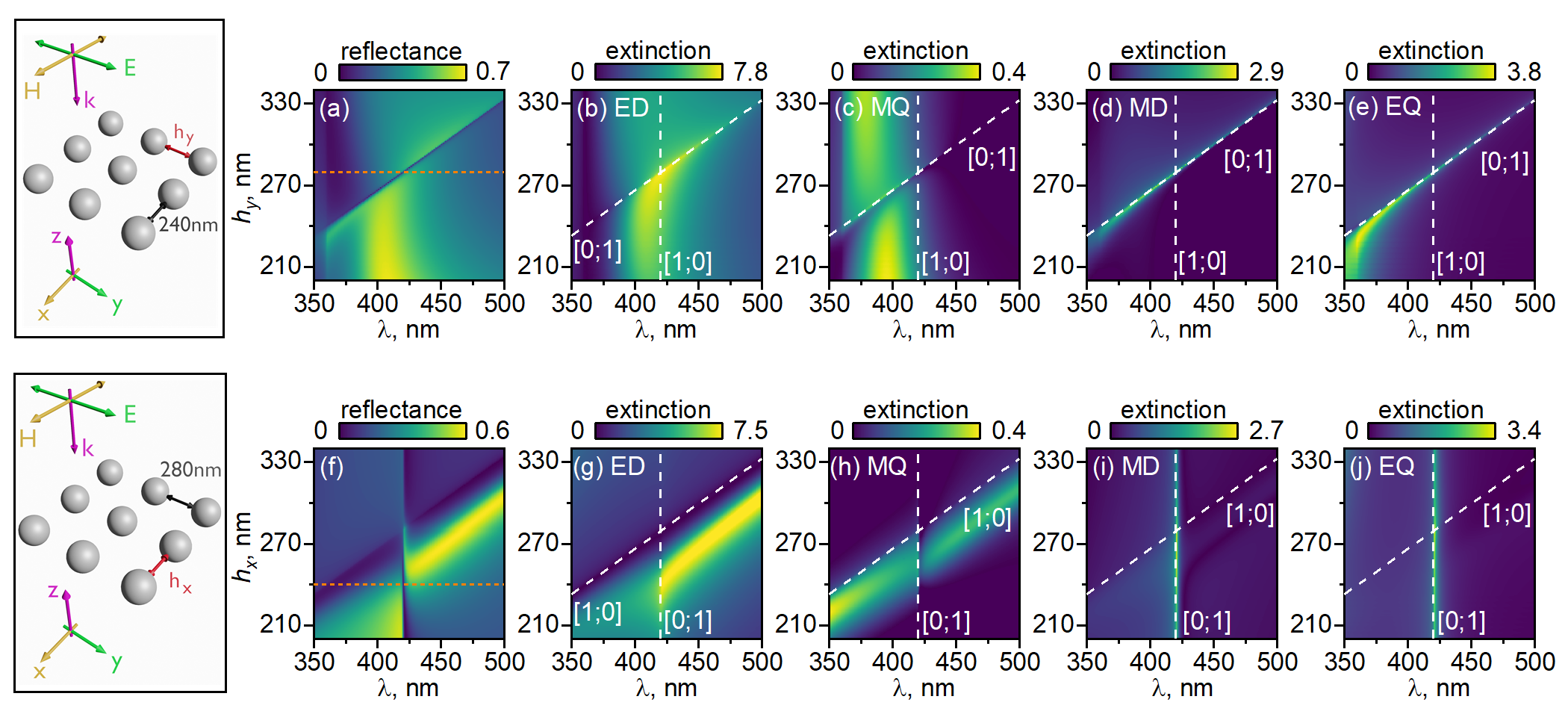}
 \caption{\label{fig:krk_all}
 (a),(f) Reflectance spectra, and (b)--(e), (g)--(j) multipole (ED, MQ, MD, EQ) decomposition of extinction efficiency for arrays with fixed $h_x =240$~nm and different $h_y$ (top), and with fixed $h_y = 280$~nm and different $h_x$ (bottom).
 Al NPs with radius $R=60$~nm have been considered in all cases.
 Notice a suppression of reflection which follows [0;1] RA, $\lambda_{\rm RA} = \sqrt{\varepsilon_h} h_y$ (see Figure~\ref{fig:field}a for details).
 The order of the multipoles appearance is chosen for the sake of consistency with eq.~\eqref{eq:qkerker}.
 Tabulated values of Al permittivity from ref.~\citenum{Smith1997TheAluminum} are used in simulations.
 }
\end{figure*}

\section{Theory}
The lattice Kerker effect in regular arrays of NPs can be understood as follows.
Consider a plane wave with frequency $\omega$ and a wavevector $|{\bf k}|=k = \sqrt{\varepsilon_h}\omega/c$ normally incident on a regular infinite array of NPs embedded in a homogeneous medium with permittivity $\varepsilon_h$ (see the sketch in Figure~\ref{fig:krk_all}).
In this case, each NP becomes indistinguishable from each other in terms of induced $\ell$-th multipoles.
The dependence of the Mie coefficients on the particle radius shows that the amplitude of the magnetic quadrupole is relatively small and higher-order modes can be ignored. The latter gives one the possibility to  restrict the discussion by $\ell=1$ and $\ell=2$ fundamental modes, i.e., electric dipole (ED), ${\bf d}$, magnetic dipole (MD), ${\bf m}$, electric quadrupole (EQ), ${\bf D}$, and magnetic quadrupole (MQ), ${\bf M}$.

Electric field in the far field reads as~\cite{landau_v2}
\begin{equation}
 \begin{split}
 {\bf E}_{sc}=-\frac{\omega^2}{c^2r}\bigg(&{\bf [[d\times n]\times n]-[m\times n]} -  \\  &\frac{i\omega}{6c} \left( {\bf [[D\times n]\times n]-[M\times n]} \right)  \bigg)e^{ikr} \ .
 \end{split}
\label{eq:multipole}
\end{equation}
Here $r$ is the distance to the observation point and
${\bf n}$ is the unit vector pointing to the observation point.
Cartesian components of EQ and MQ are $D_a=D_{ab}n_b$, $M_a=M_{ab}n_b$ (summation over repeating indices is implied), where $a,b = x,y,z$.
In a particular case of incident electric field polarized along $y$ axis, $d_x$, $m_y$, $D_{xz}$ and $M_{yz}$ are all equal zero. 
Thus, after introducing shorthands $d=d_y$, $m=m_x$, $D=D_{yz}$, and $M=M_{xz}$, the reflected field is
\begin{equation}
E_{\rm ref}=\frac{\omega^2}{c^2r}\left[ d - m + \frac{i\omega}{6c} \left( - D + M \right) \right]e^{ikr} \ .
\label{eq:Er}
\end{equation}
Dipole and quadrupole moments from the equation above can be found as: $d = \tilde{\alpha}_d E_{\rm inc}$, $m = \tilde{\alpha}_m H_{\rm inc}$, $D = \tilde{\alpha}_D \nabla_z E_{\rm inc}$, $M = \tilde{\alpha}_M \nabla_z H_{\rm inc}$, where $\tilde{\alpha}$ are the respective effective polarizabilities which depend on the geometry of the array:

\begin{equation}
\begin{split}
\tilde{\alpha}_d = i\dfrac{3\varepsilon_h}{2k^3} \tilde{a}_{1} \ , \; \tilde{\alpha}_M = i \dfrac{15\varepsilon_h^{3/2}}{k^5}\tilde{b}_{2} \ , \; \\
\tilde{\alpha}_m = i\dfrac{3\varepsilon_h}{2k^3} \tilde{b}_{1} \ , \; \tilde{\alpha}_D = i \dfrac{15\varepsilon_h^{3/2}}{k^5}\tilde{a}_{2} \ ,
\label{eq:plrzb}
\end{split}
\end{equation}
where $\tilde{a}_{\ell}$ and $\tilde{b}_{\ell}$ are expansion coefficients which take into account the interaction between NPs (not to be confused with the expansion coefficients for a single NP).
We emphasize that effective polarizabilities in the equation above depend on the properties of the individual constituent and on the lattice geometry, and also capture cross-interactions between dipoles and quadrupoles~\cite{Babicheva2019}.

For a planewave considered here, $H_{\rm inc}=\varepsilon_h E_{\rm inc}$ and $\nabla_z E_{\rm inc}=ik E_{\rm inc}$, thus eq.~\eqref{eq:Er} can be rephrased as

\begin{equation}
E_{\rm ref}=\frac{\omega^2}{c^2r}\left[\tilde{\alpha}_d-\tilde{\alpha}_m + \frac{k\omega}{6c} \left( -\tilde{\alpha}_D+\tilde{\alpha}_M \right)  \right]E_0 e^{ikr} \ .
\label{eq:kerker}
\end{equation}
It can be seen from eq.~\eqref{eq:kerker} that the reflection is  suppressed if the expression in square brackets is zero.
Thus, by appropriately tailoring both NPs properties and their arrangement, the lattice Kerker effect may emerge.

We notice that our approach allows for a comprehensive understanding of the lattice Kerker effect.
The only reported results for a plasmonic arrays~\cite{Babicheva2018c} were obtained by the finite element method (CST Microwave Studio), without a multipole decomposition of the NP lattice spectra.

\section{Results}

We demonstrate the lattice Kerker effect in regular arrays of Al NPs embedded in a homogeneous environment with $\varepsilon_h=2.25$.
In real experiment such structure corresponds to NPs deposited onto a glass substrate and subsequently covered with PMMA layer~\cite{Khlopin2017}.
It is of critical importance to match refractive indices of substrate and interparticle host medium (superstrate), to have CLRs not vanished~\cite{Auguie2010}.
NPs are arranged in infinite regular 2D lattice with periods $h_x$ and $h_y$.
The array is illuminated from the top by the plane wave with normal incidence along $z$ axis and polarization along $y$ axis. 
Narrowband suppression of the reflection can be observed in Figure~\ref{fig:krk_all}a,f.
We elaborate on this observation by plotting the contributions of each $\ell$-th mode (electric and magnetic) to the reflection in Figure~\ref{fig:krk_all}b--e,g--j (higher-order $\ell\geq 3$ modes have been taken into account in Figure~\ref{fig:krk_all}a,f, but are negligible).
To do so, we consider the extinction efficiency which is proportional only to the real part of the expression in square brackets in eq.~\eqref{eq:kerker}.
We justify this choice as follows.
On using the equations for polarizabilities~\eqref{eq:plrzb} and recalling that extinction efficiencies for electric and magnetic $\ell$-th mode are proportional to real parts of the expansion coefficients, $Q^{e}_{{\rm ext};\ell} \propto (2\ell + 1) \Re(\tilde{a}_{\ell})$ and $Q^{m}_{{\rm ext};\ell} \propto (2\ell + 1) \Re(\tilde{b}_{\ell})$, the Kerker condition can be reformulated as:
\begin{equation}
    Q^{e}_{{\rm ext};1} + Q^{m}_{{\rm ext};2} -  Q^{m}_{{\rm ext};1} - Q^{e}_{{\rm ext};2} = 0 \ .
    \label{eq:qkerker}
\end{equation}
Keeping in mind that both real and imaginary parts of the expression in square brackets in eq.~\eqref{eq:kerker} are anticipated to be zero for a case of suppressed backscattering, we shall limit the discussion only to a real part captured by the extinction efficiency, eq.~\eqref{eq:qkerker}.
The procedure for calculating extinction efficiency corresponding to each $\ell$-th mode from the equation above is described in the Appendix A.

\begin{figure}
  \includegraphics{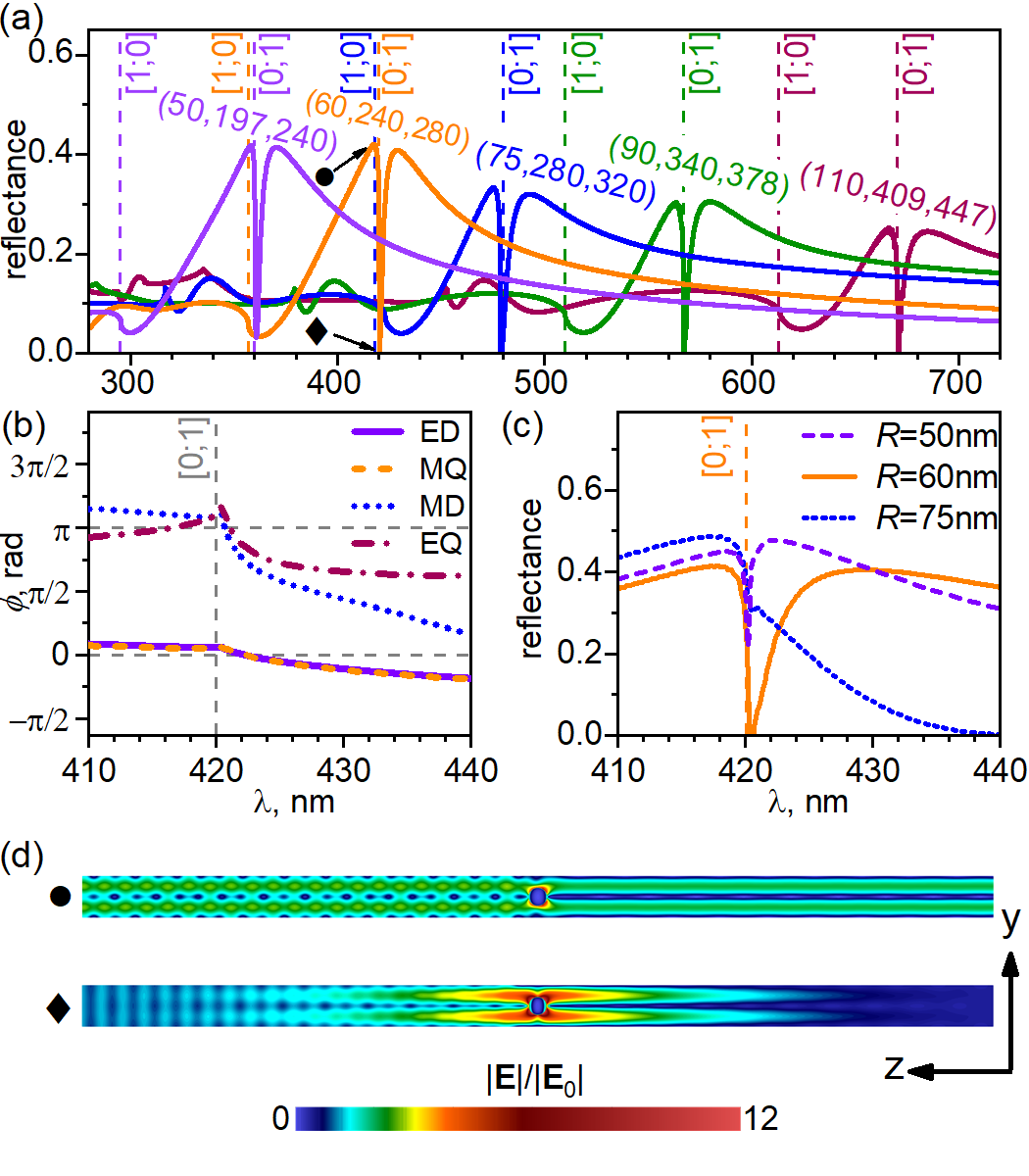}
 \caption{\label{fig:field}
(a) Reflectance for arrays with different geometrical parameters $(R, h_x, h_y)$ as marked in the legend.
Vertical dashed lines show respective spectral positions of  $[1;0]$ and $[0;1]$ RAs for each array.
(b) Multipole decomposition of the phase of the reflection amplitude $\phi={\rm arg}(E_{\rm ref})$ (eq.~\eqref{eq:Er}) for NPs array with $(60, 240, 280)$~nm (cf. orange line in (a)).
Notice $\Delta\phi=\pi$ phase difference between EQ and MD contributions on one hand, and ED and MQ counterparts on the other hand, at $\lambda = \sqrt{\varepsilon_h}h_y$, the wavelength of the lattice Kerker effect. (c) Reflectance for arrays with different NP radii $R$ and fixed $h_x=240$~nm, $h_y=280$~nm.
(d)~Electric field distribution in the $ZY$ plane for $(60, 240, 280)$~nm array at $\lambda=417.6$~nm (top, maximum reflectance) and $\lambda=420.5$~nm (bottom, zero reflectance). 
Notice rapidly vanishing  reflected and transmitted fields at $\lambda=420.5$~nm in the far zone due to strong localization of the electric field in the vicinity of NP, and high absorption shown in Figure~\ref{fig:abs}.
The dynamics of the electromagnetic field propagation at both wavelengths is demonstrated in Supporting Movie 1.
}
\end{figure}

\begin{figure}
 \includegraphics{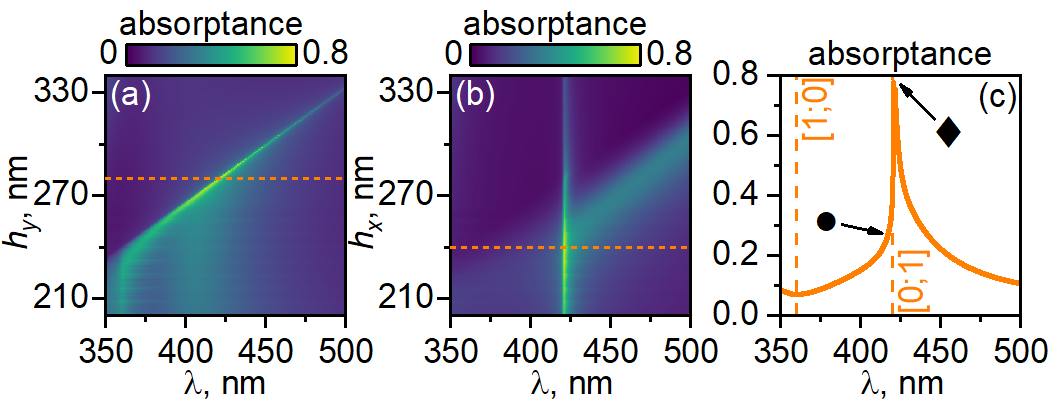}
 \caption{
 Absorption spectra for arrays with the same parameters as in Figure~\ref{fig:krk_all}, i.e., for varying (a)~$h_y$ and (b)~$h_x$, and (c) for array with $h_x=240$~nm and $h_y=280$~nm. Dashed horizontal lines indicate the lattice  periods at which the absorption spectral line $\lambda=420.5$~nm (c) is observed.}
 \label{fig:abs}
\end{figure}

Figures~\ref{fig:krk_all}b--e,g--j show that spectral properties of 2D arrays are tailored by varying one of the periods while keeping the other one constant~\cite{Li2018a,Zakomirnyi19JOSAB}.
In particular, the variation of $h_x$ keeping $h_y={\rm const}$ yields in control of $[1;0]$ Rayleigh anomaly (RA) $\lambda_{\rm RA} = \sqrt{\varepsilon_h} h_x$ and adjusts the position of ED and MQ modes.
The variation of $h_y$ and $h_x={\rm const}$ adjusts $[0;1]$ RA: $\lambda_{\rm RA} = \sqrt{\varepsilon_h} h_y$ which controls the MD and EQ modes.
Thus, Figure~\ref{fig:krk_all}b--e clearly shows that ED and MQ resonances are coupled to $[1;0]$ RA, while MD and EQ are coupled to $[0;1]$ RA.
Noteworthy, cross-interaction between different modes~\cite{Zakomirnyi19OL,Babicheva2019} results in the emergence of additional minima and maxima in Figure~\ref{fig:krk_all}b--e,g--j.
Full suppression of reflection occurs when the spectral position of resonances corresponding to different anomalies coincide with each other and the total contribution of MD and EQ modes is equal to the contribution of ED and MQ modes.

This effect is clearly visible in Fig.~\ref{fig:field}b, which shows the phases of the reflected wave created by individual multipoles. It can be seen from this figure that at $\lambda=420$~nm, phases of the ED and MQ are close to zero  while the phases of the EQ and MD are near to $\pi$. This means that the multipoles are in antiphase and destructively interfere with each other at the wavelength of the Kerker effect. It should be noted that this condition is not fulfilled near the $[1;0]$ RA.

It can be clearly seen in Figure~\ref{fig:field}a that the reflection is completely suppressed at wavelengths close to $[0;1]$ RA.
Moreover, lattice Kerker effect can be also achieved for NPs with different $R$ arranged in $2D$ lattices with properly chosen $h_x$ and $h_y$.
Thus, a complete suppression of backscattering can be tailored across UV and visible wavelength ranges, as shown in Figure~\ref{fig:field}a.
As it was shown in our recent paper~\cite[Fig.~6]{Ershov2020}, the increase of the NP radius is accompanied by the long-wavelength shift of ED, MD, EQ, MQ modes with their simultaneous broadening.  Similar behavior of the modes should be expected for the array configurations in Fig \ref{fig:field}a. This dependence is shown in Fig.~\ref{fig:field}c for arrays with different NP radii $R$ and $h_x=240$~nm, $h_y=280$~nm. It can be seen that with increase of the NP radius, the suppressed reflectance range shifts to the long-wavelength range and  broadens. 
To get more insights, we plot in Figure~\ref{fig:field}d the amplitude of electric field in $ZY$ plane for the unit cell of the array with $h_x=240$~nm, $h_y=280$~nm, at $\lambda = 420.5$ which corresponds to zero reflection.
For comparison, we also show the respective electric field distribution at $\lambda = 417.5$~nm, which corresponds to the maximum reflectance.
In the case of a maximum reflection, the standing wave originates from the interference between incident and reflected wave of comparable amplitudes.
For a zero reflectance, a symmetrical field distribution around a particle without an interference is observed. 
In this case, the amplitude of the far field rapidly vanishes along $Z$ axis according to eq.~\eqref{eq:Er}, in contrast to the case of a standing wave at $\lambda = 417.5$~nm.
Thus, at $\lambda = 420.5$, i.e., in the lattice Kerker regime, we are left only with the incident field in the far zone in the backscattering direction.

Noteworthy, the electric field enhancement near NP is significantly larger for the case of suppressed reflection due to overlapping of the ED, EQ, MD and MQ modes at the corresponding wavelength.
Thus, Kerker effect in plasmonic arrays is accompanied by high absorption of the electromagnetic energy shown in Figure~\ref{fig:abs}, up to 80\%.
The latter feature is generally not observed in weakly absorbing dielectric structures~\cite{Babicheva2017}, except few cases for amorphous~\cite{Yang2018} or crystalline~\cite{Tian2020} silicon.

\section{Summary}

To conclude, we have demonstrated the lattice Kerker effect in plasmonic arrays of Al nanoparticles, whereas for \textit{single} lossy NPs it is in principle impossible to achieve within a framework of dipole approximation~\cite{Olmos-Trigo2020}.
The plasmonic lattice Kerker effect is based on the interference suppression of dominant ED radiation (with negligible MQ impact) by the cumulative contribution of the fields produced by MD and EQ that is introduced in classical electrodynamics~\cite{landau_v2}.
We show that a complete suppression of the backscattering can be tuned within the UV and visible spectral ranges by varying geometry of arrays, i.e. radius of NPs and the distance between them.
High absorption and strong electric field localization are observed at the frequency which corresponds to the lattice Kerker effect.
The latter property is of critical importance for ultra-narrowband absorption~\cite{Li2014} and surface-enhanced Raman spectroscopy~\cite{Jha2012,Sharma2016}.

The lattice Kerker effect can also be observed in other metals besides aluminum under strict adherence to a number of conditions. 
To note, Ref.~\cite{Babicheva2018c} provides additional evidence of the manifestation of the Kerker effect in plasmonic materials. 
In particular, it was shown in Ref.~\cite{Babicheva2018c} that regular arrays of Au NPs also demonstrate resonant suppression of reflection at the border of the IR range at $\lambda = 750$~nm. Obviously, the effect can be also predicted in the IR range using large particles ($R>100$~nm)  and with a large lattice period. In this sense, gold differs just a little in comparison to all-dielectric materials which manifest the Kerker effect only in the IR spectral range.

Utilization of aluminum completely eliminates spectral restrictions and makes it possible to create conditions for the manifestation of the Kerker effect both in the entire visible and in the long-wavelength part of the UV range of the spectrum. Aluminum opens the possibility of employing the Kerker effect, particularly in spectroscopy of biomaterials with characteristic absorption bands in UV spectral range. 
This effect can also be used for photocatalysis~\cite{Chen2017} or in studies of organic and biological systems~~\cite{Tanabe2017} that exhibit strong UV absorption.
It is argued that our results expand the list of materials for fabrication of photonic devices for short-wavelength spectral  range and have academic  interest by its own as one of the first theoretical prediction of the Kerker effect in plasmonic materials.

Arrays of Al NPs also provide the possibility of selective suppression of intense lines of undesirable radiation.
For example, selective suppression of background radiation at the excitation laser wavelength is important for studies of luminescence spectra or Raman spectra~\cite{Long2002}. 
This problem can be solved by using of arrays of Al NPs as optical filters  using the backscattering suppression within one high-Q spectral line. 

In other applications, for example, related to  masking, it is important to expand the spectral range in which the suppression of radiation reflection from a surface due to the Kerker effect can be observed.

\begin{acknowledgments}
The research was supported by the Ministry of Science and High Education of Russian Federation, project no. FSRZ-2020-0008, A. E. thanks the grant of the President of the Russian Federation, agreement No. 075--15--2019--676.
\end{acknowledgments}

\bibliography{references}

\appendix
\section{Multipole field decomposition}

The total field {\bf E} and the reflectance spectra are calculated with commercial Finite-Difference Time-Domain (FDTD) package~\cite{lumerical}.
Standard approach is used to mimic infinite 2D periodic structures~\cite{Khlopin2017,Zakomirnyi17APL,Ershov2020,Utyushev2020}: periodic boundary conditions have been applied at the lateral boundaries of the simulation box, while perfectly matched layer (PML) boundary conditions were used on the remaining top and bottom sides.
Reflectance is calculated at the top of the simulation box. 
An adaptive mesh is used to reproduce accurately the nanosphere shape. 
Multipole decomposition of the extinction efficiency from Eq.~\eqref{eq:qkerker} is calculated from the spatial electromagnetic field distribution as described in ref.~\citenum{PhysRevB.84.235429}.
Namely,
The electric dipole moment is defined by the simple equation \cite{landau_v2}
\begin{equation}
    {\bf d}=\iiint_V {\bf P}d^3{\bf r}
\end{equation}
The magnetic dipole moment in the general case in the time representation reads as follows:
\begin{equation}
    {\bf m}=\int_V {\bf M}dV+\frac{1}{2c}\int_V \left[{\bf r}\times \frac{\partial {\bf P}}{\partial t}\right]dV
\end{equation}
here $M$ is magnetization. Since in the optics for the most materials  magnetic susceptibility is small~\cite{landau_v8}, the first term in the right side of the equation can be ignored. Thus, in the complex representation we obtain an expression for the magnetic dipole moment:
\begin{equation}
{\bf m}= i\frac{k}{2} \iiint_V [{\bf P} \times{\bf r} ] d^3{\bf r}
\end{equation}
Similar expressions can be obtained for quadrupole moments:
\begin{equation}
\begin{split}
    {\bf \hat{D}}&=3\iiint_V \left({\bf P}\otimes {\bf r} +{\bf r}\otimes {\bf P}\right) d^3{\bf r},\\
    {\bf \hat{M}}&=\frac{2}{3} ik\iiint_V {\bf r}\otimes\left[{\bf P}\times {\bf r}\right]d^3{\bf r},
\end{split}
\end{equation}
here $\varepsilon$ is permittivity of the particles, ${\bf P}=(\varepsilon-1)/4\pi{\bf E}$ is polarization density, $k$ is the wavenumber.
The corresponding extinction cross sections read as follows: 

\begin{equation}
\begin{split}
    \sigma_{ext}^d &= \frac{4\pi k}{\sqrt{\varepsilon_h}|{\bf E}_0|^2}\Im( {\bf E}_0^*\cdot{\bf d}),\\
    \sigma_{ext}^m &= \frac{4\pi k}{\sqrt{\varepsilon_h}|{\bf H}_0|^2}\Im( {\bf H}_0^*\cdot{\bf m}),\\
    \sigma_{ext}^D &= -\frac{\pi k}{3\sqrt{\varepsilon_h}|{\bf E}_0|^2}\Im\left[ (\nabla \otimes{\bf E}_0^*+{\bf E}_0^* \otimes\nabla):\hat{D}\right],\\
    \sigma_{ext}^M &= -\frac{2\pi k}{\sqrt{\varepsilon_h}|{\bf H}_0|^2}\Im\left[ (\nabla\otimes {\bf H}_0^*+{\bf H}_0^* \otimes\nabla):\hat{M}\right],\\
        \end{split}
\end{equation}
here ${\bf E}_0$ and ${\bf H}_0$ are components of the incident electromagnetic field.
The respective extinction efficiency $Q$ is defined as the extinction cross section $\sigma$ normalized to the geometric cross section $\pi R^2$ of a single particle with radius $R$. 

\end{document}